\documentclass[reprint,twocolumn]{revtex4}

\usepackage{graphicx}
\usepackage{dcolumn}
\usepackage{bm}
\usepackage{color}
\usepackage{amsmath,amsthm}
\usepackage{amssymb,bm}
\usepackage{mathrsfs}
\usepackage{comment}
\usepackage{ulem}

\begin{document}

\title{Precise determination of excitation energies in condensed-phase molecular systems based on exciton-polariton measurements}
\author{Nguyen Thanh Phuc}
\author{Akihito Ishizaki}
\affiliation{Department of Theoretical and Computational Molecular Science, Institute for Molecular Science, Okazaki 444-8585, Japan}
\affiliation{Department of Structural Molecular Science, The Graduate University for Advanced Studies, Okazaki 444-8585, Japan}

\begin{abstract}
The precise determination of the excitation energies in condensed-phase molecular systems is important for understanding system-environment interactions as well as for the prerequisite input data of theoretical models used to study the dynamics of the system. The excitation energies are usually determined by fitting of the measured optical spectra that contain broad and unresolved peaks as a result of the thermally random dynamics of the environment. Herein, we propose a method for precise energy determination by strongly coupling the molecular system to an optical cavity and measuring the energy of the resulting polariton. The effect of thermal fluctuations induced by the environment on the polariton is also investigated, from which a power scaling law relating the polariton's linewidth to the number of molecules is obtained. The power exponent gives important information about the environmental dynamics.
\end{abstract}


\maketitle

Embedded in a high density of environmental particles, the excitation energies of molecules in condensed phase can be modified from their values in gas phase by the static influence of various kinds of system-environment interactions including electrostatic interaction and hydrogen bonding~\cite{Eccles83, Damjanovic02, Spangler77, Sturgis96, Sturgis97, Witt02}, as well as the effects of molecular conformation~\cite{Warshel87, Nowak90}. Therefore, a precise determination of the excitation energies of condensed-phase molecular systems is significant for the understanding of system-environment interactions. Moreover, the energy values are prerequisite as input data for almost all theoretical models used to study the dynamics of molecular systems~\cite{Adolphs06, Ishizaki09}.
The excitation energies of condensed-phase molecular systems are usually determined by fitting of the measured optical spectra. The optical spectra, however, often contain broad and unresolved peaks as a result of the thermally random dynamics of the environment interacting with the molecular system. Moreover, to extract excitation energy information from the optical spectra, it is necessary to develop a theory of optical spectra that addresses the often sophisticated spectral density of the environment. A variety of approximations are sometimes used to reduce the complexity of the calculations~\cite{Vulto98, Wendling02, Adolphs06}. Consequently, it is desirable to develop an alternative approach that can precisely determine the excitation energies of condensed-phase molecular systems without requiring detailed information about the environmental random dynamics. 

In this Letter, we propose a method for the precise determination of the excitation energies of condensed-phase molecular systems by strongly coupling the molecules to an optical cavity and measuring the energy of the polariton that results from the hybridization of the degrees of freedom of light and matter. Strong coupling of molecules to an optical cavity has already been realized in many experimental platforms~\cite{Ebbesen16, Lidzey98, Lidzey99, Lidzey00, Hobson02, Tischler05, Holmes07, Cohen08, Bellessa14, Schwartz11, Cohen13, Mazzeo14, Cacciola14, Gambino15, George15, Long15, Muallem16, George15b, Shalabney15, Saurabh16, Chikkaraddy16}. It has led to a variety of interesting phenomena and important applications including the control of chemical reactivity~\cite{Hutchison12, Simpkins15, Herrera16, Galego16, Thomas16, Thomas19, Hiura19, Lather19}, enhancement of transport~\cite{Hutchison13, Andrew00, Feist15, Schachenmayer15, Orgiu15, Zhou16}, nonlinear optical properties of organic semiconductors with applications to optoelectronic devices~\cite{Herrera14, Bennett16, Kowalewski16, Kowalewski16b}, and polariton lasing and condensate~\cite{Cohen10, Cwik14, Lerario17, Plumhof14}.
The underlying mechanism that allows a precise determination of the excitation energies of condensed-phase molecular systems is that the polariton appears as a sharp peak in optical spectrum under the inflence of strong coupling between the cavity mode and the electronic excitations of molecules inside the cavity.
This is related to the effect of vibronic or polaron decoupling found in the Holstein-Tavis-Cummings (HTC) model that describes molecules with a single vibrational mode that are coupled to an optical cavity~\cite{Spano15, Herrera16, Wu16, Zeb18, Herrera18} and the extended model~\cite{Pino18}. However, since the polariton is a collective superposition of a large number of electronic excitations of molecules [see Eq.~\eqref{eq: polariton state}] and therefore can be vulnerable to decoherence, the effect of thermal fluctuations induced by the environment on the polariton state at finite temperatures, which is not captured in the HTC model and its extension, is a nontrivial and important issue. 

By investigating the effect of thermal fluctuations induced by the environment on the polariton, a power scaling law relating the polariton's linewidth to the number of molecules coupled to the cavity is determined. The power exponent strongly depends on the environment's dynamic time scale. As such, information on environmental dynamics can be extracted from the polariton spectrum obtained for a variable number of molecules. 
Since the polariton contains both light and matter degrees of freedom, its energy can be obtained by either cavity-transmission or molecular-absorption spectroscopies. In the latter, the polariton needs to be in a bright state with respect to molecular absorption. However, this condition is not satisfied if there are pairs of molecules with the opposite orientations, such as in the case of random orientations. The effective Rabi frequencies for an ensemble of identical molecules or molecular complexes with random orientations are derived. The molecular-absorption and cavity-transmission spectroscopies are calculated for several cases of molecular systems with different types of orientations, in which the potential of using polariton for precise determination of excitation energies in condensed-phase molecular systems is demonstrated.

\begin{figure}[htbp] 
  \centering
  \includegraphics[keepaspectratio]{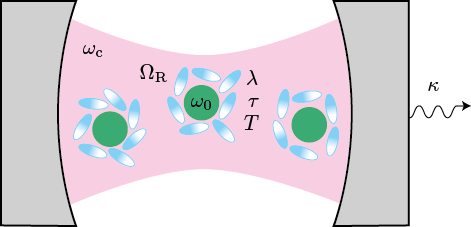}
  \caption{Schematic illustration of a system of condensed-phase molecules coupled to a single mode of an optical cavity (light magenta) with frequency $\omega_\mathrm{c}$. Each molecule (green sphere) with the electronic excitation energy $\hbar\omega_0$ interacts with an independent surrounding environment represented by blue ellipsoids. The thermal dynamics of the environments at a finite temperature $T$ induces energy fluctuations in the molecules that are characterized by the reorganization energy $\lambda$ and the relaxation time scale $\tau$. The coupling strength between the cavity and a molecule is given by the single-emitter Rabi frequency $\Omega_\mathrm{R}$. Due to the finite transmitivity of the cavity mirrors (grey plates), the cavity photon has a loss rate of $\kappa$.}
  \label{fig: system}
\end{figure}

\textit{Effect of environmental thermal fluctuations on the polariton's linewidth.--}
We consider a system of $N$ identical molecules whose electronic excitations are coupled to a single mode of an optical cavity (Fig.~\ref{fig: system}) via dipole interaction
\begin{align}
\hat{H}_\mathrm{mc}=\frac{\hbar\Omega_\mathrm{R}}{2}\sum_{m=1}^N 
\left(|\mathrm{e}_m\rangle\langle \mathrm{g}_m|\hat{a}+|\mathrm{g}_m\rangle\langle \mathrm{e}_m|\hat{a}^\dagger\right),
\end{align} 
where $\Omega_\mathrm{R}$ is the so-called single-emitter Rabi frequency that characterizes the coupling strength between the cavity and a molecule, $\hat{a}$ denotes the annihilation operator of the cavity photon, and $|\mathrm{g}_m\rangle$ and $|\mathrm{e}_m\rangle$ represent the electronic ground and excited states, respectively, of the $m$th molecule. In this case, we assume that all molecules inside the cavity have the same orientation such that their Rabi couplings are equal. Molecules with different orientations will be considered later. 

Each molecule in condensed phase is assumed to be coupled to an independent environment, which is modeled by an ensemble of harmonic oscillators $\hat{H}_\mathrm{e}=\sum_{m=1}^N \sum_\xi \hbar \omega_{m,\xi}\hat{b}_{m,\xi}^\dagger \hat{b}_{m,\xi}$, where $\omega_{m,\xi}$ and $\hat{b}_{m,\xi}$ represent the frequency and annihilation operator of the $\xi$ mode of the environment surrounding the $m$th molecule. The dynamics of the environment at a finite temperature induces energy fluctuations in the electronic excited states of the molecules as given by the Hamiltonian~\cite{May-book}
\begin{align}
\hat{H}_\mathrm{me}=\sum_{m=1}^N 
\left[ \hbar\omega_0+\sum_{\xi}g_{m,\xi}\left(\hat{b}_{m,\xi}^\dagger+\hat{b}_{m,\xi}\right)\right]
|\mathrm{e}_m\rangle\langle \mathrm{e}_m|.
\label{eq: molecule-environment interaction}
\end{align}
Here, $\hbar\omega_0$ is the molecule's excitation energy and $g_{m,\xi}$ denotes the coupling strength between the $m$th molecule and the $\xi$ mode of the environment. The dynamics of the environment is characterized by the relaxation function $\Psi_m(t)=(2/\pi)\int_0^\infty \text{d}\omega J_m(\omega)\cos(\omega t)/\omega$, where $J_m(\omega)=\pi\sum_\xi g_{m,\xi}^2\delta(\omega-\omega_{m,\xi})$ is the spectral density. When the spectral density is given by the Drude-Lorentz form, $J_m(\omega)=2\lambda_m\tau_m\omega/(\tau_m^2\omega^2+1)$, the relaxation function has an exponential form, $\Psi_m(t)=2\lambda_m\exp(-t/\tau_m)$, where $\lambda_m$ is the environmental reorganization energy, which is usually employed to characterize the system-environment coupling strength, and $\tau_m$ is the characteristic timescale of the environmental relaxation or reorganization process~\cite{Ishizaki10}. The time evolution of the system's reduced density operator can be solved in a numerically accurate fashion using the hierarchical equations of motion approach for example~\cite{Tanimura06}.

The molecular absorption spectrum can be expressed in terms of the system's dynamical quantities as~\cite{Mukamel-book}
\begin{align}
\mathcal{A}(\omega)=\text{Im}
\left\{\frac{i}{\hbar} \int_0^\infty \text{d}t e^{i\omega t} \text{Tr}\left[\hat{\mu}\mathcal{G}(t)\hat{\mu}^\times\hat{\rho}_0\right]\right\},
\end{align}
where $\hat{\mu}=\sum_{m=1}^N \left(\mu_m |\mathrm{e}_m\rangle\langle\mathrm{g}_m| + \mu_m^* |\mathrm{g}_m\rangle\langle\mathrm{e}_m|\right)$ is the total transition dipole moment operator with $\mu_m$ being the matrix element of the transition dipole moment of the $m$th molecule, and $\hat{\mu}^\times\hat{\rho}\equiv \hat{\mu}\hat{\rho}-\hat{\rho}\hat{\mu}$. Here, the density operator $\hat{\rho}_0=|\mathrm{G}\rangle \langle \mathrm{G}|$ with $|\mathrm{G}\rangle=\prod_{m=1}^N |\mathrm{g}_m\rangle \otimes |0\rangle_\mathrm{c}$ being the ground state of the cavity-molecule system in which all the molecules are in their electronic ground states and the cavity is in the vacuum state. The superoperator $\mathcal{G}(t)$ describes the time evolution of the system in Liouville space. In the following numerical demonstration, we set the parameters of the molecular system and the environment to be $\omega_0=12400\,\text{cm}^{-1}$, $\lambda=50\,\text{cm}^{-1}$, $\tau=100$ fs, and $T=300$ K, which are typical values in photosynthetic pigment-protein complexes~\cite{Adolphs06, Ishizaki09}. For simplicity, we assume that the parameters of the environments are equal. The cavity frequency is taken to be $\omega_\mathrm{c}=12450\,\text{cm}^{-1}$, i.e., with a detuning of $50\,\text{cm}^{-1}$ from the molecule's excitation energy. The cavity's $Q$-factor ($\kappa=\omega_\mathrm{c}/Q$) is set to be $Q=10^4$. When many molecules are coupled to a single mode of the optical cavity, the collective Rabi frequency $\sqrt{N}\Omega_\mathrm{R}$ rather than the single-emitter Rabi frequency $\Omega_\mathrm{R}$ determines the polariton energy. 

To investigate the effect of thermal fluctuation due to the environment on the polariton's linewidth, we calculate the full width at half maximum of the lower-polariton peak, which represents the polariton with an energy less than both the molecule's excitation energy and the cavity frequency in the molecular-absorption spectrum. It is determined that in the absence of molecule-cavity coupling, the molecular absorption peak has a broad linewidth of approximately $291\,\text{cm}^{-1}$, which is larger than the typical separation between absorption peaks of different molecules~\cite{Adolphs06}. In contrast, when the molecules are coupled to the cavity mode, the polariton peak in the absorption spectrum has a much smaller linewidth that decreases with an increase of the number $N$ of molecules coupled to the cavity.
The $N$-dependence of the polariton's linewidth is determined to follow a power scaling law $\Delta\nu_\mathrm{LP}=\Delta_0 N^{-\alpha}$ with $\Delta_0= 138\pm4\,\text{cm}^{-1}$ and $\alpha= 0.57\pm 0.02$ (for the collective Rabi frequency $\sqrt{N}\Omega_\mathrm{R}=0.1\,\mathrm{eV}$) obtained via the linear fitting of numerical data in the logarithmic scale~\cite{Supp}. 

To obtain physical insight into the effect of environmental thermal fluctuations on the polariton's linewidth , we first assume that the thermal fluctuations do not alter the structure of the lower polariton, which has the form of
\begin{align}
|\mathrm{LP}\rangle=c_1\prod_{m=1}^N |\mathrm{g}_m\rangle\otimes|1\rangle_\mathrm{c}
+c_2|\mathrm{B}\rangle\otimes|0\rangle_\mathrm{c},
\label{eq: polariton state}
\end{align}
where $c_1$ and $c_2$ are coefficients of the superposition that satisfy $|c_1|^2+|c_2|^2=1$. Here, $|n\rangle_\mathrm{c}$ denotes the Fock state with $n$ cavity photons and $|\mathrm{B}\rangle =(1/\sqrt{N})\sum_{m=1}^N |\mathrm{e}_m\rangle$ is the so-called bright state, which is a superposition of electronic excitations of all molecules coupled to the cavity. 
Given that each of the molecular electronic excitations $|\mathrm{e}_m\rangle$ is coupled to the environmental modes via the interaction given by Eq.~\eqref{eq: molecule-environment interaction}, the lower polariton $|\mathrm{LP}\rangle$ with the aforementioned structure can be regarded as having an effective interaction in which the number of modes increases by a factor of $N$ but the coupling strength to each mode decreases by a factor of $1/N$. As such, the effective spectral density $J_\mathrm{LP}(\omega)$ and the reorganization energy $\lambda_\mathrm{LP}$ for the lower polariton are modified by a factor of $1/N$ because the spectral density is proportional to the mode density and the coupling strength squared.

The effect of the environment on the linewidth of the lower polariton in the absorption spectrum depends on the environmental dynamics. In the inhomogeneous broadening limit $\sqrt{k_\mathrm{B}T\lambda_\mathrm{LP}}\gg \tau^{-1}$, which corresponds to the slow environmental dynamics, the lineshape has a Gaussian form with the linewidth given by $\sqrt{k_\mathrm{B}T\lambda_\mathrm{LP}}$~\cite{May-book}. As a result, the polariton's linewidth should follow an $N^{-1/2}$ scaling. In the opposite limit of homogeneous broadening $\sqrt{k_\mathrm{B}T\lambda_\mathrm{LP}}\ll \tau^{-1}$, which corresponds to the fast environmental dynamics, the lineshape has a Lorentzian form with the linewidth given by $k_\mathrm{B}T\lambda_\mathrm{LP}\tau$. As a result, the polariton's linewidth should follow as $N^{-1}$ scaling.

\begin{figure}[tbp] 
  \centering
  \includegraphics[keepaspectratio]{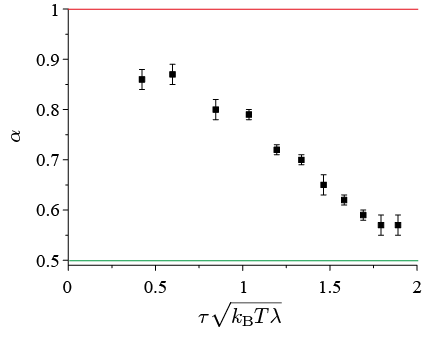}
  \caption{Dependence of the power exponent $\alpha$ in the power scaling $\Delta\nu_\mathrm{LP}\propto N^{-\alpha}$ of the lower polariton's linewidth with respect to the number of molecules coupled to the cavity on the dimensionless quantity $\tau\sqrt{k_\mathrm{B}T\lambda}$ that characterizes the environment's dynamic motion. Here, $\lambda$ is the reorganization energy, $\tau$ is the relaxation time scale, and $T$ is the temperature of the environment ($k_\mathrm{B}$ is the Boltzmann constant). The red (green) line indicates the value of $\alpha=1$ ($\alpha=0.5$), which is the expected value of $\alpha$ in the inhomogeneous (homogeneous) broadening limit $\tau\sqrt{k_\mathrm{B}T\lambda}\ll 1$ ($\tau\sqrt{k_\mathrm{B}T\lambda}\gg 1$) under the assumption that the environmental thermal fluctuations do not affect the structure of the lower polariton.}
  \label{fig: power exponent}
\end{figure}

To examine the validity of the preceding analysis, in which it was assumed that the thermal fluctuations do not affect the structure of the lower polariton, we numerically calculate the $N$-dependence of the polariton's linewidth in the absorption spectrum for different values of the reorganization energy $\lambda$. It was determined that the power scaling law of $\Delta\nu_\mathrm{LP}\propto N^{-\alpha}$ is well-satisfied with a strong $\lambda$-dependence of the power exponent as shown in Fig.~\ref{fig: power exponent}. Consequently, environmental dynamics information can be extracted from the power exponent $\alpha$ obtained by measuring the lower polariton energy for a variable number of molecules coupled to the cavity. It is evident that $\alpha$ generally decreases with an increase of $\lambda$ and seems to approach a steady value close to $\alpha=1$ ($\alpha=0.5$) in the inhomogeneous (homogeneous) broadening limit. The remaining deviation should, however, be attributed to the effect of environmental thermal fluctuations on the structure of the lower polariton.

\textit{Polariton energy.--}
The energy of the lower polariton can be obtained by diagonalizing the Hamiltonian of the cavity-molecule system, yielding
\begin{align}
\omega_\mathrm{LP}=\frac{1}{2}\left[\omega_\mathrm{c}+\omega_0-\sqrt{(\omega_\mathrm{c}-\omega_0)^2+N\Omega_\mathrm{R}^2}\right].
\end{align}
For a sufficiently large Rabi frequency, $\sqrt{N}\Omega_\mathrm{R}\gg |\omega_\mathrm{c}-\omega_0|$, it reduces to $\omega_\mathrm{LP}\simeq (\omega_\mathrm{c}+\omega_0-\sqrt{N}\Omega_\mathrm{R})/2$. Therefore, by repeating the measurement of the lower polariton energy with variable molecular density, in which $N$ is varied, or with variable number of photons in the cavity, by which $\Omega_\mathrm{R}$ is varied, we can obtain the molecular excitation energy $\hbar\omega_0$ via a simple linear fitting, given that the cavity frequency $\omega_\mathrm{c}$ is known, for example, based on the transmission spectroscopy measurement of the bare cavity.

The deviation $|\nu_\mathrm{LP}-\omega_\mathrm{LP}|$ of the position of the lower-polariton peak in the absorption spectrum from its energy as a function of $N$ was investigated. It was determined that the deviation increases with an increase of the number of molecules coupled to the cavity prior to saturation, following the function $|\nu_\mathrm{LP}-\omega_\mathrm{LP}|=A-Be^{-\gamma N}$~\cite{Supp}.
Using the exponential fitting procedure, we determine the saturation value of $|\nu_\mathrm{LP}-\omega_\mathrm{LP}|$ to be $A\simeq 27\,\mathrm{cm}^{-1}$, which is smaller than $\Delta\nu_\mathrm{LP}\simeq 55\,\mathrm{cm}^{-1}$ for $N=5$ (for the same collective Rabi frequency $\sqrt{N}\Omega_R=0.1\,\mathrm{eV}$). 
The dependence of $\Delta\nu_\mathrm{LP}$ and $|\nu_\mathrm{LP}-\omega_\mathrm{LP}|$ on the Rabi frequency $\Omega_\mathrm{R}$ was also investigated. It was determined that $\Delta\nu_\mathrm{LP}$ decreases with an increase of $\Omega_\mathrm{R}$ before it saturates for a sufficiently large collective  Rabi frequency~\cite{Supp}. The saturation value is determined by the number of molecules coupled to the cavity. The deviation $|\nu_\mathrm{LP}-\omega_\mathrm{LP}|$ follows a power scaling $|\nu_\mathrm{LP}-\omega_\mathrm{LP}|=C\Omega_\mathrm{R}^{-\eta}$~\cite{Supp}.

Next, we consider a more general system of molecular complex composed of molecules with different excitation energies and transition dipole moments. 
Both the magnitude and the sign of the Rabi coupling can differ from one molecule to another in the system. 
In the following numerical demonstration, we consider a system of $N=3$ molecules with excitation energies $\omega_1=12400\,\text{cm}^{-1}$, $\omega_2=12500 \,\text{cm}^{-1}$, and $\omega_3=12600 \,\text{cm}^{-1}$. The Rabi frequencies associated with the three molecules are taken to be $\sqrt{N}\Omega_\mathrm{R}^{(1)}=0.1$ eV, $\Omega_\mathrm{R}^{(2)}=\Omega_\mathrm{R}^{(1)}\sqrt{3}/2$, and $\Omega_\mathrm{R}^{(3)}=-\Omega_\mathrm{R}^{(1)}/\sqrt{2}$. We also consider coupling between electronic excitations of different molecules given by the Hamiltonian
\begin{align}
\hat{H}_\mathrm{mm}=\sum_{m\not=n}
\hbar V_{mn}|\mathrm{e}_n\rangle\langle\mathrm{g}_n|\otimes |\mathrm{g}_m\rangle\langle\mathrm{e}_m|,
\end{align}
where the coupling matrix elements $V_{mn}$ between the $m$th and the $n$th molecules satisfy $V_{mn}=V_{nm}^*$. In this case we take $V_{mn}=50 \,\text{cm}^{-1}$. 

There is a relatively sharp and isolated peak in the molecular absorption spectrum that corresponds to the lower polariton~\cite{Supp}. The linewidth of the peak is determined to be $\Delta\nu_\mathrm{LP}\simeq 85\,\mathrm{cm}^{-1}$, which has a comparable magnitude to the linewidth of the polariton peak in the investigated case of $N=3$ identical molecules.
Besides the lower-polariton peak, there is a broad peak that contains the absorption spectra of the upper-polariton as well as two remaining energy eigenstates. In the case of identical molecules, these two energy eigenstates are dark states with respect to molecular absorption and thus do not appear in the absorption spectrum. Due to the difference in the excitation energy and the Rabi coupling between the molecules, as well as the inter-molecular electronic couplings, they are no longer fully dark states. However, given that these eigenstates consist mainly of the degrees of freedom of matter, their linewidths are broad compared with those of the polaritons. 

The energy of the lower polariton is obtained by diagonalizing the Hamiltonian of the molecule-cavity system, which in this case is a $4\times 4$ matrix
\begin{align}
\begin{pmatrix}
\omega_\mathrm{c} & \Omega_\mathrm{R}^{(1)}/2 & \Omega_\mathrm{R}^{(2)}/2 & \Omega_\mathrm{R}^{(3)}/2 \\
\Omega_\mathrm{R}^{(1)}/2 & \omega_1 & V_{12} & V_{13} \\
\Omega_\mathrm{R}^{(2)}/2 & V_{21} &\omega_2 & V_{23}\\
\Omega_\mathrm{R}^{(3)}/2 & V_{31}& V_{32} & \omega_3
\end{pmatrix}.
\end{align}
The deviation of the position of the lower-polariton peak in the absorption spectrum from the lower polariton energy was determined to be $|\nu_\mathrm{LP}-\omega_\mathrm{LP}|\simeq 19\,\mathrm{cm}^{-1}$, which is smaller than $\Delta\nu_\mathrm{LP}$. By repeating the measurement of the lower-polariton energy with variable molecular density and/or variable cavity frequency, for example, by adjusting the distance between two mirrors and using the genetic algorithm for a multivariable fitting~\cite{Kinnebrock94, Pohlheim99, Bruggemann04, Adolphs06}, the excitation energies of the molecules can be determined or at least the accuracy of their values obtained using other approaches can be evaluated.

In the case of an ensemble of identical molecules with different orientations coupled to a single mode of an optical cavity, the energy of the polariton can be obtained, as in the case of one molecule, by using an effective Rabi frequency
$\Omega_\mathrm{R}^\mathrm{eff}=\sqrt{\sum_{m=1}^N \left|\Omega_\mathrm{R}^{(m)}\right|^2}$,
where $\Omega_\mathrm{R}^{(m)}$ is the Rabi coupling associated with the $m$th molecule. If the orientation of the molecules are random, using $\langle\cos^2\theta\rangle_\theta=1/2$, we obtain $\Omega_\mathrm{R}^\mathrm{eff}=\Omega_\mathrm{R}\sqrt{N/2}$ with $\Omega_\mathrm{R}$ being the Rabi frequency of one molecule.

Similarly, if an ensemble of identical molecular complexes with different orientations is coupled to a single mode of an optical cavity, the energy of the polariton can be obtained, as in the case of one molecular complex, by using effective Rabi frequencies for energy eigenstates (excitons) of the molecular complex
$\Omega_\mathrm{R}^{i, \mathrm{eff}}=\sqrt{\sum_{m=1}^N \left|\Omega_\mathrm{R}^{i, (m)}\right|^2}$.
Here $i$ represents the exciton energy eigenstates of each molecular complex, and $\Omega_\mathrm{R}^{i, (m)}$ represents the Rabi coupling associated with the $i$th exciton in the $m$th molecular complex. 
If the orientation of the molecular complexes is random, the effective Rabi frequencies reduce to $\Omega_\mathrm{R}^{i, \mathrm{eff}}=\Omega_\mathrm{R}^i \sqrt{N/2}$ with $\Omega_\mathrm{R}^i$ being the Rabi frequency of the $i$the exciton in one molecular complex.

\textit{Cavity transmission spectrum.--}
We have demonstrated that the sharp and isolated peak of the lower polariton appears in the molecular absorption spectrum, which can be used for precise determination of the excitation energies of molecules. However, if there are pairs of identical molecules with opposite orientations such that their Rabi couplings have the same magnitude but the opposite signs, the polariton state would become a dark state with respect to molecular absorption~\cite{Supp}. This situation is encountered especially in a system of identical molecules or identical molecular complexes with random orientations. In this case, given that the polariton always involves the degrees of freedom of the cavity, its energy can be obtained from cavity transmission spectroscopy measurements.

For a numerical demonstration of the cavity transmission spectrum, we consider a system of $N=4$ identical molecules that form two pairs of molecules with opposite orientation. As a result, the Rabi couplings satisfy $\Omega_\mathrm{R}^{(2)}=-\Omega_\mathrm{R}^{(1)}$ and $\Omega_\mathrm{R}^{(4)}=-\Omega_\mathrm{R}^{(3)}$. In this case, we take $\sqrt{N}\Omega_\mathrm{R}^{(1)}=0.1$ eV and $\Omega_\mathrm{R}^{(3)}=\Omega_\mathrm{R}^{(1)}\sqrt{3}/2$. The other parameters of the system and the environment are the same as those of the aforementioned system that was investigated. 
There is a relatively sharp peak of the lower polariton with a linewidth of $\Delta\nu_\mathrm{LP}\simeq 67\,\mathrm{cm}^{-1}$, which has a comparable magnitude to that of the molecular-absorption spectrum of a system of $N=4$ molecules with the same orientation~\cite{Supp}. There is also a very small and flat transmission spectrum at approximately $12400\,\mathrm{cm}^{-1}$ due to the three energy eigenstates of the system that consists mainly of the degrees of freedom of matter. In the absence of thermal fluctuation from the environment, these energy eigenstates do not appear in the cavity transmission spectrum. Their signals in the cavity transmission spectrum should therefore be attributed to the thermal fluctuation of the environment, which affects the structures of the energy eigenstates by inducing small mixing of the degrees of freedom of light and matter.

\textit{Conclusions.--}
We have demonstrated that a precise determination of the excitation energies of condensed-phase molecular systems is possible by strongly coupling the molecules to an optical cavity and measuring the energy of the polariton, which is a mixture of light and matter degrees of freedom. 
The polariton's linewidth is determined to exhibit a power scaling with respect to the number of molecules coupled to the cavity mode. The power exponent strongly depends on the environment's dynamic time. Therefore, the environmental dynamics information can be extracted from the polariton spectrum measured for a variable number of molecules.
The exciton-polariton-based approach proposed here is the first step in the development of new methods for precise measurement and/or control of various physical properties of condensed-phase molecular systems~\cite{Phuc18, Phuc19}, which is significant from the perspective of both fundamental science and technological application.

\begin{acknowledgements}
This work was supported by JSPS KAKENHI Grant Numbers 19K14638 (N.~T.~Phuc), 17H02946 and 18H01937, and JSPS KAKENHI Grant Number 17H06437 in Innovative Areas ``Innovations for Light-Energy Conversion (I$^4$LEC)'' (A.~Ishizaki). 
\end{acknowledgements}


\end{document}